
\input harvmac
\noblackbox
%
%

\def\hf{{1\over2}}

\def\ajou#1&#2(#3){\ \sl#1\bf#2\rm(19#3)}
\def\apm{{\alpha^\prime}}

\def\frac#1#2{{#1 \over #2}}

\def\etp{e^{2\phi}}
\def\eps{\epsilon}
\def\mn{{\mu\nu}}
\def\ls{{\lambda\sigma}}

\def\vx{{\vec x}}

\def\p{\partial}

\def\to{{\rightarrow}}
\def\WL{P{\rm exp} \left( \int_0^{2 \pi R} d x^4 A_4 \right) }

\hyphenation{Mor-ris-on}
%
%
\lref\duff{M. Duff, Class. Quantum Grav. {\bf 5} (1988) 189. }
\lref\HL{J. A. Harvey and J. Liu, Phys. Lett. {\bf B268} (1991) 40.}
\lref\banks{T. Banks, M. Dine, H. Dijkstra and W. Fischler,
Phys. Lett. {\bf B212} (1988) 45.}
\lref\gibb{G. W. Gibbons and P. J. Ruback, Commun. Math. Phys. {\bf 115}
(1988) 267.}
\lref\sch{N. S. Manton  and B. J.Schroers, Ann. Phys. {\bf 225} (1993) 290.}
\lref\dab{A. Dabholkar and J. A. Harvey, Phys. Rev. Lett.  {\bf 63} (1989) 719
\semi
A. Dabholkar, G. Gibbons, J. A. Harvey, and F. Ruiz Ruiz,
Nucl. Phys. B {\bf B340} (1990) 33.}
\lref\sena{A. Sen, preprint TIFR/TH/94-03 hep-th/9402002.}
\lref\senb{A. Sen, preprint TIFR-TH-94-08 hep-th/9402032.}
\lref\GNO{P. Goddard, J. Nuyts and D. Olive, Nucl. Phys. {\bf B125}
(1977) 1.}
\lref\WO {E. Witten and D. Olive, Phys. Lett. {\bf 78B} (1978) 97. }
\lref\CSF{ E. Cremmer, J. Scherk and S.  Ferrara,
Phys. Lett. {\bf 74B} (1978) 61.}
\lref\GHL{J. P. Gauntlett, J. A. Harvey and J. Liu,
Nucl. Phys. {\bf B409} (1993) 363.}
\lref\harr{B. Harrington and H. Shepard, Phys. Rev. {\bf D17} (1978) 2122.}
\lref\rossi{P. Rossi, Nucl. Phys. {\bf B149} (1979) 170.}
\lref\hetsol{A. Strominger, Nucl. Phys. {\bf B343} (1990) 167;
E: Nucl. Phys. {\bf B353} (1991) 565.}
\lref\MO{C. Montonen and D. Olive, Phys. Lett. {\bf 72B} (1977) 117. }
\lref\osborn{H. Osborn, Phys. Lett. {\bf 83B} (1979) 321.}
\lref\gpy{D. J. Gross, R. D. Pisarski and L. G. Yaffe, Rev. Mod. Phys.
{\bf 53} (1981) 43.}
\lref\fromzum{B. Zumino, Phys. Lett. {\bf B69}, (1977) 369.}
\lref\don{
J. A. Harvey and A. Strominger, Comm. Math. Phys. {\bf 151} (1993) 221. }
\lref\smon{J. P. Gauntlett,
Nucl. Phys. {\bf B400} (1993) 103; Nucl. Phys. {\bf B411} (1993) 443.}
\lref\blum{J. Blum, preprint EFI-94-04 (hep-th/9401133).}
\lref\grossman{B. Grossman, Phys. Lett. {\bf A61} (1977) 86.}
\lref\dualrefa{A. Font, L. E. Ib\'a\~nez, D. L\"ust and F. Quevedo,
Phys. Lett. {\bf B249} (1990) 35;
S. J. Rey, Phys. Rev. {\bf D43}  (1991) 526.}
\lref\dualrefb{
A. Sen, Nucl. Phys. {\bf B404} (1993) 109; Phys. Lett. {\bf B303} (1993) 22;
Mod. Phys. Lett. {\bf A8} (1993) 2023;
J. H. Schwarz and A. Sen, Nucl. Phys. {\bf B411}
(1994) 35.}
\lref\inprep{J. P. Gauntlett and J. A. Harvey, work in progress}
\lref\HM{J. Horne and G. Moore, ``Chaotic Coupling Constants,'' preprint
YCTP-P2-94, RU-94-25 (hep-th/9403058)}
\lref\witten{E. Witten, Nucl. Phys. {\bf B202} (1982) 253.}
\lref\orbi{L. Dixon, J. A. Harvey, C. Vafa and E. Witten, Nucl. Phys. {\bf
B261} (1985) 678;
Nucl. Phys. {\bf B274} (1986) 285. }
\lref\tod{A. Todorov, Inv. Math. {\bf 61} (1980) 25.}
\lref\nati{N. Seiberg, Nucl. Phys. {\bf B303} (1988) 286.}
\lref\rohmw{R. Rohm and E. Witten, Ann. Phys. {\bf 170} (1986) 454.}
 \lref\bddf{T. Banks, M. Dine, H. Dijkstra and W. Fischler,
Phys. Lett. {\bf B212} (1988) 45.}
\lref\sgrav{E. Cremmer, J. Scherk and S. Ferrara, Phys. Lett. {\bf 74B} (1978)
61 \semi
E. Bergshoeff, M. de Roo and B. de Wit, Nucl. Phys. {\bf B182} (1981) 173 \semi
M. K. Gaillard and B. Zumino, Nucl. Phys. {\bf B193} (1981) 221 \semi
M. de Roo, Nucl. Phys. {\bf B255} (1985) 515.}
\lref\kkref{R. Sorkin, Phys. Rev. Lett. {\bf 51} (1983) 87 \semi
D. Gross and M. Perry, Nucl. Phys. {\bf B226} (1983) 29.}
\lref\narain{K. S. Narain, Phys. Lett. {\bf B169} (1986) 41.}
\lref\nsw{K. S. Narain, M. H. Sarmadi, and E. Witten, Nucl. Phys. {\bf B279}
(1987) 369.}
\lref\stw{A. Shapere, S. Trivedi and F. Wilczek, Mod. Phys. Lett. {\bf A6}
(1991) 2677.}
\lref\jm{C. V. Johnson and R. C. Myers, preprint hep-th/9406069. }
\lref\atiyah{M. F. Atiyah, ``Geometry of Yang-Mills Fields'' in Lezioni
Fermiane, Accademia Nazionale dei Lincei Scuola Normale Superiore,
Pisa (1979). }
\lref\adhm{M. F. Atiyah, N. J. Hitchin, V. G. Drinfeld and Yu. I. Manin,
Phys. Lett. {\bf 65A} (1978) 425.}
\lref\ah{M.F. Atiyah and N. J. Hitchin, The Geometry and Dynamics of
Magnetic Monopoles, Princeton University Press, 1988.}
\lref\thooft{G. 't Hooft, Nucl. Phys. {\bf B153} (1979) 141 ; Acta Phys.
Austriaca Suppl.
{\bf XXII} (1980) 531.}
\lref\wittwist{E. Witten, Nucl. Phys. {\bf B202} (1982) 253.}
\lref\hofftwo{ G. 't Hooft, Commun. Math. Phys. {\bf 81} (1981) 267.}
\lref\modrefs{This result is implicit in the work of G. 't Hooft, Phys. Rev.
{\bf D14} (1976) 3432,
see also C. Bernard, Phys. Rev. {\bf D19} (1979) 3013. }
\lref\unpub{J. P. Gauntlett and J. A. Harvey, unpublished.}
\lref\nfourref{S. Mandelstam, Nucl. Phys. {\bf B213} (1983) 149 \semi
L. Brink, O. Lindgren and B. Nilsson, Phys. Lett. {\bf 123B}(1983) 323 \semi
P. S. Howe, K. S. Stelle and P. K. Townsend, Nucl. Phys. {\bf B214} (1983) 519
\semi
P. S. Howe, K. S. Stelle and P. K. Townsend, Phys. Lett. {\bf 124B} (1983) 55
\semi
E. Martinec, Phys. Lett. {\bf B171} (1986) 189.}
\lref\wbrane{C. Callan, J. A. Harvey and A. Strominger, Nucl. Phys. {\bf B359}
(1991) 40.}
\lref\dufflu{M. Duff and J. Lu, Nucl. Phys, {\bf B354} (1991) 129; {\bf B357}
(1991) 129.}
\lref\dufflutwo{M. Duff and J. Lu, Phys. Rev. Lett. {\bf 66} (1991) 1402.}
\lref\bmt{P. J. Braam, A. Maciocia, and A. Todorov, Invent. math. {\bf 108}
(1992) 419.}
\lref\kmir{F. Bogomolov and P. J. Braam,
Commun. Math. Phys. {\bf 143} (1992) 641.}
\lref\hoffmod{G. 't Hooft, Phys. Rev. {\bf D14} (1976) 3432.}
\lref\dist{J. Distler and S. Kachru, PUPT-1464, hep-th@xxx/9406091.}
\lref\bhrefs{recent refs on bh, Giddings, Polchinski, Strominger,
Kallosh, Duff???, what else?}
\lref\coleman{S. Coleman, ``Classical Lumps and their Quantum Descendants'' in
Aspects of Symmetry, Cambridge University Press, 1985.}
\lref\moore{G. Moore, hep-th/9305139.}
\lref\gir{L. Girardello, A. Giveon, M. Porrati, and A. Zaffroni,
hep-th/9406128.}
%
%
\Title{\vbox{\baselineskip12pt
\hbox{EFI-94-36}
\hbox{hep-th/9407111}}}
{\vbox{\centerline{$S$-Duality and the Spectrum of Magnetic }
\centerline{Monopoles in Heterotic String Theory}}}
{
\baselineskip=12pt
\bigskip
\centerline{Jerome P. Gauntlett$^*$ and Jeffrey A. Harvey}
\bigskip
\centerline{\sl Enrico Fermi Institute, University of Chicago}
\centerline{\sl 5640 Ellis Avenue, Chicago, IL 60637 }
\centerline{\it Internet: jerome@yukawa.uchicago.edu}
\centerline{\it Internet: harvey@poincare.uchicago.edu}

\bigskip
\medskip
\centerline{\bf Abstract}
We discuss the predictions of $S$-duality for the monopole
spectrum of four-dimensional heterotic string theory resulting from
toroidal compactification. We discuss in detail the spectrum of
 ``$H$-monopoles" , states that are magnetically charged with respect to the
$U(1)$ groups arising from the dimensional reduction of the ten dimensional
antisymmetric tensor field.
Using an assumption concerning the correct treatment of collective
coordinates in string theory we find results which are consistent with
$S$-duality.
\vskip 1.8in
\noindent
${}^*$Address after September 1 1994:
California Institute of Technology, Pasadena, CA 91125

}


\Date{7/94}
%

\newsec{Introduction}
It has long been hoped that four-dimensional theories might be found which
exhibit a duality that interchanges weak and strong coupling, thus extending
the remarkable features of special two-dimensional theories  to four
dimensions.
The discovery of such a duality would certainly
have profound consequences.

There is circumstantial evidence for such a duality
in N=4 super Yang-Mills theory spontaneously broken down to a subgroup
containing a $U(1)$ factor and hence magnetic monopoles.
This can be traced back to the GNO conjecture
concerning a dual  theory  of magnetic monopoles  \GNO\ and
the conjecture of Montonen and Olive \MO\ that there exist
a dual gauge theory in which the coupling $g^2 \to 1/g^2$ and electric
and magnetic charges are exchanged.
Some evidence for this conjecture
comes from the existence of a duality
invariant formula for the mass of states in terms of
their magnetic and electric charges. This formula must be exact
for states which furnish a sixteen-dimensional (short) representation
of the $N=4$ supersymmetry algebra \WO. These states include
the massive vector multiplet, the photon multiplet and,
after quantization
of the fermion zero modes about the monopole solution, the
monopole supermultiplet
\osborn.

A natural extension of the duality conjecture occurs when the effects of
a non-zero theta parameter, $\theta$,  are considered.   The duality which
acts on the gauge coupling as $g^2 \rightarrow 1/g^2$ is then naturally
extended to a
$SL(2,Z)$ group of transformations acting on $\lambda = \theta / 2 \pi +
i/g^2$.
It was noted by Sen \sena\ that this extension makes new non-trivial
predictions which do not follow from the symmetry $g^2 \rightarrow 1/g^2$
alone. In particular, given the relatively
weak assumption that a state with electric
charge one and magnetic charge zero (i.e. a massive $W^+$ state) exists
in the ${\bf 16}$ of $N=4$ supersymmetry for all
$\lambda$, one deduces the existence of an infinite tower of
stable states with electric and magnetic
charges determined by a pair of relatively
prime integers $(p,r)$,
each in the ${\bf 16}$ of $N=4$.
In a  semi-classical  analysis these states
should arise from
the existence of
specific normalizable harmonic forms on the reduced moduli space
of classical monopole solutions.
In a recent remarkable paper
one of these predictions, the existence of an unique normalizable
anti-self-dual harmonic two-form
on the two-monopole moduli space, was born out \senb\ (see also \refs{\gibb,
\sch}).

There has also been recent discussion of the possibility of
$SL(2,Z)$ duality (
``$S$-duality'' ) in
toroidally compactified heterotic
string theory. Indications for such a duality come from
many different sources. These include the mysterious non-compact
symmetries of dimensionally reduced supergravities and in particular the
$SL(2,R)$ symmetry which is characteristic
of four-dimensional $N=4$ supergravity \sgrav, the soliton-like
behavior of fundamental strings \dab, a geometrical duality between
strings and fivebranes \duff, the existence of fivebrane solutions
to string theory  and the suggestion of a weak-strong coupling
duality involving strings and fivebranes \refs{\hetsol,\wbrane, \dufflu}.
On the basis of such ideas $S$-duality was proposed in general
four-dimensional heterotic strings \dualrefa, but the clearest
evidence at present comes from the study of
toroidal compactifications \dualrefb.
Since $N=4$ super Yang-Mills is embedded in
the four dimensional low-energy field theory limit of this
compactification,
the $S$-duality
conjecture generalizes the $S$-duality in the $N=4$ super Yang-Mills theory.

The precise group which should be
involved in $S$-duality is not completely clear. At the classical
level one finds an  $SL(2,R)$ symmetry. It is believed that this should
be broken to $SL(2,Z)$ by conventional field theory instantons
\refs{\dualrefa,\stw}.
It is
possible that other backgrounds or more stringy effects break
it to an even smaller subgroup.  The possibility of the full $SL(2,Z)$ acting
is intimately connected with the fact that $E_8 \times E_8$ and
$\rm{Spin}(32/Z_2)$ are self-dual gauge groups in the sense of  ref. \GNO.
In this paper we will assume that
the full $SL(2,Z)$ should act, but for the most part we will only
discuss the implications of the $Z_2$ subgroup which takes weak
to strong coupling at $\theta=0$.

In attempting to extend $S$-duality from $N=4$ super-Yang-Mills theory to
string theory
one finds new qualitative
features.
One of these features is that the coupling constant and the $\theta$
angle become dynamical fields. Thus $S$-duality acts on the combination
\eqn\axiodil{\lambda = \Psi + i e^{-\Phi} }
as
\eqn\sact{\lambda \rightarrow {a \lambda + b \over c \lambda +d}}
where  $\Psi$ is the axion field defined by
\eqn\psidef{H = - e^{2 \Phi} {}^* d \Psi ,}
$\Phi$ is the string
dilaton field, and $ad-bc=1$ with $a,b,c,d \in Z$.
This, plus the fact
that $\Psi$ changes by $1$ in encircling a fundamental string suggests
that $S$-duality should be regarded as a discrete gauge symmetry in
string theory \sena.

The study of $S$-duality in string theory is also
complicated by the existence of several different types of
magnetic monopoles. At generic points in the
moduli space of Narain compactification ,
 ${\cal M}_N$ \refs{\narain,\nsw} ,
the low-energy gauge
group is $U(1)^{28}$ with $16$ of the $U(1)$ factors arising from the
Cartan subalgebra of $E_8 \times E_8$, $6$ arising from the
off-diagonal components of the metric $g_{\mu n}$,
and $6$
arising from the off-diagonal components of the anti-symmetric
tensor field $B_{\mu n}$ ($\mu,\nu=0,\dots,3; m,n=4,\dots,9$).
Labelling the $U(1)$ field strengths
as $F_{\mu \nu}^a$, $a=1 \ldots 28$,  the action of $S$-duality
is
\eqn\fsdual{ F_{\mu \nu}^a \rightarrow (c \lambda_1 +d)F_{\mu \nu}^a
              + c \lambda_2 (ML)_{ab} {\tilde F}_{\mu \nu}^b }
where $M$ is a $28 \times 28$ matrix of scalars (Narain moduli)
and $L$ is the $O(6,22)$ invariant metric. The scalars $M$, the
Einstein metric and the fermions are all left invariant by $S$-duality.

The magnetic monopole solutions predicted by $S$-duality will thus
include essentially all previously studied types of magnetic monopoles,
that is 't Hooft-Polyakov or BPS
monopoles,
Kaluza-Klein monopoles \kkref, and $H$-monopoles \refs{\rohmw, \bddf, \GHL}.
In addition,
as one moves in ${\cal M}_N$ (i.e. by varying the asymptotic
values for the scalars $M$) the various $U(1)$ factors mix with each other
so at a generic point the monopoles of the
diagonal
$U(1)$'s will be a linear combination of these various types.
Furthermore these monopoles are only
light compared to the string scale in some regions in
${\cal M}_N$. Thus in general gravitational and string corrections to the
solutions must be included.

In recent work Sen has given a comprehensive review of the evidence for
$S$-duality in toroidally compactified heterotic string
theory and has initiated a program to test $S$-duality by finding a set of
specific
testable predictions
which can be checked at weak-coupling and in the field theory limit \sena.
The work of \senb\ verifies one of these predictions, but
strictly speaking
does not provide evidence for
$S$-duality in string theory independent of its existence
in $N=4$ super-Yang-Mills theory.
The other predictions made
in \sena\ regard the spectrum after semi-classical quantization of
$H$-monopoles.  These are magnetic monopoles which carry magnetic
charges under $U(1)$ groups which arise from dimensional reduction of the
antisymmetric tensor  field $B_{MN}$ in ten dimensions.
The purpose of this paper is to study whether the predictions of
$S$-duality
for $H$-monopoles are
also born out.  As we will see,  it does not seem possible to answer this
question
completely in the field theory limit, in spite of the fact that $H$-monopoles
can be made arbitrarily light.  However with one crucial assumption
concerning the correct treatment of collective coordinates in string theory,
at least one of these predictions
is true. Since the $H$-monopole solutions have a non-trivial dependence on
the compactified
dimensions the duality predictions concern the full ten-dimensional
low-energy field theory. Since  the low-energy field
theory is presumably not
a consistent quantum theory on its own,  our results should be regarded as
evidence for duality of the full string theory, or at least its
toroidal compactification,
if our
assumption concerning collective coordinates is correct.
Precisely what duality should
mean for the full string theory is far from clear,  at the end of this paper
we will make some speculative remarks on this subject.

\newsec{Bogomol'nyi States in Heterotic String Theory}

Since $S$-duality involves the interchange of strong and weak coupling,
to test
whether it is a valid symmetry of string theory it is necessary, at least
with our current techniques, to consider quantities that are exact at tree
level
and check  their invariance under the duality. As discussed
in \sena, in the context of six dimensional toroidal compactifications
their are several such quantities, including the mass spectrum of
``Bogomol'nyi states", states
in the {\bf 16} of the N=4 supersymmetry algebra.
These representations only exist for
states whose mass saturates a Bogomol'nyi bound
determined by their electric and magnetic
charges. Thus
the relation between mass and charge for these states
must be an exact quantum relation\WO. Since the electric and
magnetic charges are not
renormalized in theories with $N=4$
supersymmetry \nfourref\ we deduce that both
the tree level
charges and masses of Bogomol'nyi states are exact.

Thus to test $S$-duality we first determine the spectrum
of electrically charged first-quantized string states which saturate
the Bogomolnyi bound. Assuming that these Bogomol'nyi states exist at
all values of the coupling, $S$-duality  predicts the existence of
magnetic monopoles and dyons with degeneracies equal to those of the
corresponding dual states. Semi-classical reasoning at weak coupling
then translates this into properties
of monopole moduli spaces.

One rather surprising feature of heterotic string theory is the existence
of an infinite
tower of electrically charged
Bogomol'nyi states of increasing mass \dab. Any state
constructed as a tensor product of the right-moving (superstring)
oscillator ground state and an arbitrary left-moving state satisfying
the constraint
\eqn\lrequal{N_L -1 = {1 \over 2}(p_R^2 - p_L^2) }
preserves half of the spacetime supersymmetry and satisfies a Bogomol'nyi
bound
\eqn\massfor{M^2 = {1 \over 8} p_R^2}
where $(p_L, p_R) \in \Gamma_{22,6}$ are the electric charges of
the state and $\Gamma_{22,6}$ the even,
self-dual, Lorentzian lattice determining the compactification. The
partition function for these states is
\eqn\zbog{Z_{Bog}(q,\bar q) = { \Theta_\Gamma(q,\bar q) \over
                               \eta(q)^{24} } \times (8-8) }
with $\Theta_\Gamma$ the lattice theta function,
\eqn\defth{\Theta_\Gamma(q,\bar q)= \sum_{(p_L,p_R) \in \Gamma_{22,6}}
                                     q^{p_L^2 /2} {\bar q}^{p_R^2 /2} ,}
$\eta$  the  Dedekind eta function,
\eqn\etdef{\eta(q) = q^{1/24} \prod_n (1-q^n) , }
and the factor of $(8-8)$ arises from the $8$ bosonic and $8$
fermionic right-moving ground states.

In general, to test $S$-duality at a particular
point in ${\cal M}_N$ we should first expand $Z_{Bog}$ and
identify terms with equal powers of $q$ and $\bar q$. The prefactor
of such a term gives the degeneracy of states with $U(1)^{28}$ electric
charges given by $(p_L,p_R)$. Applying an $SL(2,Z)$ transformation
we predict the existence of magnetic monopole solutions with the
dual magnetic charges and the same degeneracy. To the first few orders
in $q$ this gives
\eqn\zexp{ Z_{Bog}(q,\bar q) = (8-8) q^{-1} \left( 1+ 24q +
                              (24 \cdot 25 /2 + 24)q^2
                              + \cdots \right) \sum_{(p_L,p_R)}
							   q^{p_L^2/2} {\bar q}^{p_R^2 /2}  .}
Thus at lowest order we have electrically charged states with
$p_L^2- p_R^2=  2$ with degeneracy $1$ up to the $16$ fold-degeneracy
arising from $N=4$ supersymmetry and at the next level we have
states with $p_L^2 - p_R^2 =0 $ with degeneracy $24$.

Before proceeding we should mention two possible obstacles to
carrying out this procedure. The first is the continuous spectrum of
the theory. Since the spectrum contains massless particles (graviton,
photon, dilaton, etc.), massive electrically charged states are
part of a continuum of states if we work in infinite volume. Thus
it is not clear that the degeneracy of such states is well
defined, particularly at strong coupling. We will not be able to
shed much light on this question, and some aspects of this continuous
spectrum will become evident in our quantization of the monopole
degrees of freedom. We will assume nonetheless that the degeneracy
of states can be defined and provides a test of $S$-duality. The second
obstacle is connected with gravitational effects. Since there is an
infinite tower of Bogomol'nyi states in string theory, one might expect that
for any fixed value of the string coupling these must eventually be treated
as (extremal) black holes since their mass eventually exceeds the
Planck mass\foot{Because of the presence of the dilaton
in string theory this is not necessarily the case. For example,
the BPS monopole solution in low-energy string theory does not become
a black hole even for arbitrarily large Higgs expectation value \HL,
in contrast to  Einstein-Yang-Mills-Higgs theory.}.
On the other hand the mass  of these states is
proportional to the string coupling  times the Planck mass,
so by taking the coupling to be very small we can always treat the
low-lying states as solitons rather than as black holes since their
Compton wavelength exceeds their Schwarzschild radius by a factor
of the inverse string coupling.

In principle we would like to construct magnetic monopole solutions
to string theory as conformal field theories and identify and quantize
their collective coordinates directly in string theory. Since it is
not yet known how to do this in detail, we will have to rely on
the low-energy field theory as a guide. To do
this we need to work in a limit where the low-energy field theory
is a good approximation (modulo one problem to be discussed
later). We can achieve this by
working at a point where all radii $R_i$ of the six-torus are
large compared to the string scale and where all Wilson line
expectation values which break $E_8 \times E_8 $ to $U(1)^{16}$
are small compared to the string scale. In addition we should
consider only states with masses small compared to the string scale.
For $R_i$ large this means we should only consider states with
vanishing winding number or equivalently states with a Kaluza-Klein origin.

A simple example may be useful. The Narain moduli
are the constant values of the metric, anti-symmetric
tensor field, and Cartan sub-algebra gauge fields with
indices tangent to $T^6$. We take $g_{mn} = R^2 \delta_{mn},
B_{mn}=0$ and $A_m^I \ne 0$ with $R^{-1}, A_m^I << \sqrt{\alpha'}$.
For vanishing winding number the momenta $(p_L,p_R)$ are given
by \nsw\
\eqn\narmod{\eqalign{p_L^I & = P^I  \cr
                     p_L^m & = \half g^{mn} (M_n + A_n^I P^I) \cr
					 p_R^m & = p_L^m \cr}}
with $P^I$ an element of the $E_8 \times E_8$ root lattice and
$M_n$ labels the momenta on $T^6$. Thus
the lightest states with $N_L=0$, $p_L^2 = p_R^2 +2$ and multiplicity $1$
have $M_n=0$, $P^I \ne 0$, mass squared
$M^2 = (A_n^I P^I)^2 /32 R^2$, and are charged under the $U(1)^{16}$
coming from $E_8 \times E_8$ but are neutral under the
$U(1)^{12}$ coming from the metric and antisymmetric tensor \foot{The Wilson
line $A^I_m$
leads to mixing between the $U(1)$ factors arising from the
ten-dimensional  $E_8 \times E_8$ and
metric fields. What is meant here is that the diagonal $U(1)$ comes
predominantly from $E_8 \times E_8$ for small $A^I_m$.}.
Light states with $N_L=1$, $p_L^2 = p_R^2$ and multiplicity $24$ have
$P^I =0$, $M_n \ne 0$, mass squared $M^2 = (M_n)^2 /32 R^2$ and
are charged under $U(1)^{12}$ but neutral under
$U(1)^{16}$. More precisely these states carry charge under
the $U(1)^6$ coming from the metric and are neutral under
the $U(1)^6$ coming from the antisymmetric tensor field
(since $(p_L\pm p_R)/2$ determine the metric and antisymmetric
tensor $U(1)$ charges, respectively).

Under the transformations \sact, \fsdual\ with $\Psi=0$ and
$a=d=0, c=-b=1$ these electrically charged states transform
into states with magnetic charge. The states with $p_L^2= p_R^2 +2$
are dual to magnetic monopoles of the $U(1)^{16}$, that is BPS
monopoles, while the states
with $p_L^2=p_R^2$ are dual to magnetic monopoles of the $U(1)^6$
coming from the antisymmetric tensor field, that is they transform
into $H$ monopoles\foot{That $S$-duality relates electrically charged
states with respect
to the metric $U(1)^6$ to magnetically charged states with respect
to the antisymmetric tensor $U(1)^6$ is due to the off-diagonal form
of the matrix $L$.}. The BPS monopoles are predicted to have multiplicity
one in the ${\bf 16}$ of $N=4$ supersymmetry.
In the limit considered here gravitational corrections to the
monopole moduli space should be insignificant so the spectrum
is in accord with $S$-duality as in \senb.

The $H$ monopoles on
the other hand are predicted to have multiplicity $24$ in the ${\bf 16}$
of $N=4$ and it is this prediction which we wish to test. There is
a more precise prediction given by decomposing the $24$ states
under the unbroken symmetries of the theory. We can work at a
point in ${\cal M}_N$ where the theory has a $SO(5) \times SO(3)$
symmetry with the $SO(5)$ arising as a subgroup of the global
$SO(6)$ of $N=4$ super Yang-Mills and $SO(3)$ the group of spatial rotations.
Under this symmetry group the massive ${\bf 16}$ of $N=4$ decomposes as
${\bf 16} \rightarrow (5,1)+(1,3)+(4,2)$ while the $24$ states at the
first excited left-moving level transform as $16 (1,1) + (5,1)+(1,3)$.
Note in particular that  the tensor product includes massive spin two
states.
If $S$-duality is correct we should find the same representations in
the spectrum of $H$-monopoles of charge one, that is we should
find exactly $24 \times 16$ harmonic forms on the one $H$-monopole
moduli space with the specified quantum numbers.

\newsec{The Spectrum of $H$-Monopoles}

\subsec{$H$-Monopoles}
In this section we will concentrate on the construction
of the light $H$ magnetic monopoles discussed in the previous
section.
Let us first review briefly the construction presented in \GHL.
Monopole
solutions are most easily constructed out of periodic fivebranes
i.e. by wrapping the fivebrane around the six torus \refs{\rohmw,\hetsol}.
To illustrate the construction without unnecessary complication we will
start at a point in ${\cal M}_N$ with $SU(2) \times U(1)^{27}$ symmetry
( with $SU(2) \subset E_8 \times E_8$ )
and discuss the final breaking of $SU(2)$ to $U(1)$ explicitly.
Let $x^4 \equiv x^4 + 2 \pi R$ be a coordinate
on one $S^1$ ; $x^i$, $ i=1 \cdots 3 $,
the three spatial coordinates, and $\mu, \nu, ...$
indices which run from $1$ to $4$. The solution of \GHL\  obeys the ansatz
\eqn\fivebr{\eqalign{
F_\mn &=  \pm\hf\eps_\mn{}^\ls F_{\ls}\cr
H_{\mn\lambda} &= \mp\eps_{\mn\lambda}{}^\sigma\partial_\sigma\phi\cr
g_\mn& =\etp\delta_\mn. \cr
}}
To this order, $H=dB-{\apm\over 30}\omega$ where $\omega$ is the Yang-Mills
Chern-Simons
three form and hence the Bianchi identity,
\eqn\bianch{dH=-{\apm\over 30}{\rm Tr}F\wedge F,}
determines the form of the dilaton via
\eqn\dil{\nabla_\rho\nabla^\rho\phi=-{\apm\over 60} {\rm Tr}F^2.}
The solution is thus completely determined by
a (anti-) self-dual $SU(2)$ connection on $R^3 \times S^1$.
Such solutions can be constructed from an
array of instantons on $R^4$ which is periodic
in $x^4$. If the instantons all have the same scale size and gauge orientation
then
one finds the solution
\eqn\two{ A_\mu = {\bar \Sigma}_{\mu \nu} \nabla^\nu \ln f (x) }
with
\eqn\singlem{\eqalign{f (\vx, x^4) &= 1 + \sum_{k=-\infty}^\infty
{\rho^2 \over r^2 + (x^4-x^4_0 + 2 \pi k R)^2 }\cr
&= 1 +{\rho^2\over2Rr}\sinh{r\over R}
\bigg/\left(\cosh{r\over R}-\cos{x^4-x^4_0\over R}\right)}}
where $r=| \vec x - \vec x_0|$ and $(\vec x_0,x^4_0)$ give the
location of the instanton and ${\bar \Sigma}_{\mu \nu}$ is the matrix-valued
't Hooft tensor. For $\rho$ finite the $SU(2)$ gauge fields
fall off as $1/r^3$ at large $r$ and are thus dipole rather than monopole
fields.
On the other hand, one finds that $H_{ij4} \sim \epsilon_{ijk} x^k / r^3$
indicating
that these solutions are $H$-monopoles\foot{Note that Sen's definition of
an $H$-monopole \sena\ includes some of the other moduli and gauge fields but
for the solutions we are considering
the asymptotic behavior is exactly the same as the definition being used here.}
 as required by duality.

It is known that generalizations of these solutions exist at points where the
$SU(2)$
symmetry is broken to $U(1)$, but unfortunately an explicit representation does
not seem to be available.  The solutions can be described following \gpy\ in a
somewhat heuristic manner which should be correct at large $R$ where
the multi-instanton moduli space degenerates into copies of the one-instanton
moduli space, by continuity one
would expect the same general structure at finite $R$. Start  with
an infinite string of instantons located at $x^i = x_0^i$, $x^4 = 2 \pi R n$,
$n \in Z$,
with identical scale size but with a gauge orientation which rotates by
$\omega$ between
instantons with $\omega$ in a $U(1)$ subgroup of $SU(2)$.  The resulting vector
potential will be periodic up to a gauge transformation by $\omega$ and has a
vanishing
Wilson line about $x^4$. If one then performs an improper gauge transformation
$U= \omega^{\alpha (x^4/2 \pi R)}$ one obtains a single periodic instanton
with a non-trivial Wilson line $\Omega(x^i)=\WL$ which breaks $SU(2)$ to $U(1)$
if $\alpha$ is
not an integer.
By integrating the Bianchi identity \bianch\ one concludes that the solution
is indeed an $H$-monopole with magnetic charge proportional to
the Pontryagin number.

Although we have not done so, it should be possible to construct
these solutions perturbatively in the strength of the $SU(2)$ breaking.
It is known (see e.g. \atiyah\ ) that there is a well-defined perturbative
expansion about the $5k$ parameter
't Hooft solutions which gives  local parameters
for the full $8k$ parameter moduli space described in the
ADHM construction \adhm.

Having described a classical solution with the required properties
we have to determine whether the quantum states have the structure
required by duality. As usual the low-lying states may be studied by
quantization of the bosonic and fermionic collective coordinates
corresponding to zero-energy deformations  of the monopole.
Let us first discuss the bosonic collective coordinates. At finite
$\rho$ and $\Omega=1$ from \two, \singlem, we see that the  moduli consist of
the
``center of mass'' coordinates $(\vec x_0, x^4_0)$ and the scale
size $\rho$.  In addition there are three collective coordinates
which describe the $SU(2)$ orientation of the instantons. Since
the center of $SU(2)$ acts trivially, these three collective coordinates
are properly treated as coordinates on $SU(2)/Z_2 = SO(3) = S^3/Z_2$.
With a non-trivial Wilson
line the treatment of these three collective coordinates is
somewhat different since one should only consider $SU(2)$ transformations
which
commute with the value of the Wilson line at spatial infinity.
On the other hand, spatial rotations which
cannot be undone by $SU(2)$ transformations commuting
with the Wilson line $\Omega(\infty)$ will  also then have to be
included. With trivial Wilson line one can equally well think of
the $SO(3)$ collective coordinates as arising from spatial
rotations since rotations and $SU(2)/Z_2$ gauge transformations
have equivalent actions on the instanton configuration. When
the Wilson line is non-vanishing it is simplest to think
of all three collective coordinates as arising from rotations. One again
concludes that these collective coordinates parametrize $SO(3) = S^3/Z^2$.

The bosonic collective coordinates are coordinates on the one
monopole moduli space, ${\cal M}^1$.
The metric
on ${\cal M}^1$  is determined as follows. If we let
$Z^i$ denote the collective coordinates  and $A^0_\mu(x,Z^i)$
the general monopole solution then zero modes are given
up to a gauge transformation by varying the classical solution with
respect to the $Z^i$:
\eqn\zmod{\delta_i A_\mu = \partial_i A^0_\mu -D^0_\mu \epsilon_i  , }
where the gauge parameters $\epsilon_i$ are determined by demanding that
the zero modes be orthogonal to local gauge transformations
\eqn\bkgnd{D^0_\mu \delta_i A^\mu = 0.}
The metric on ${\cal M}$ is
\eqn\mmet{{\cal G}_{ij} = \int d^3 x \int d x^4 {\rm Tr} (\delta_i A_\mu
\delta_j A_\mu ) .}
The
gauge parameter $\epsilon_i$ provides a natural connection on ${\cal M}^1$
with covariant derivative
\eqn\coder{s_i = \partial_i + [\epsilon_i,{~~}]}
and field strength
\eqn\pdef{\phi_{ij} = [s_i,s_j] . }
There is also a hyperK\"ahler structure on ${\cal M}^1$ which is induced from
the hyperK\"ahler structure
$J^{(m)}$ on $R^3 \times S^1$:
\eqn\comstr{{\cal J}^{(m)}\,_i\,^{j} = \int d^3x \int dx^4 \, {J^{(m)}\,_\mu}
\,^{\nu}\,{\rm Tr}({\delta_i}{A^\mu}
{\delta_k}{A_\nu}){\cal G}^{kj}. }

In addition to the bosonic zero modes there are fermionic zero modes
which are paired with the bosonic zero modes via the N=4 supersymmetry
\fromzum.
For each fermionic zero mode we introduce a Grassmann odd collective
coordinate. Techniques to carry out the explicit reduction of the
four dimensional action to ${\cal M}^1$ are discussed
in \refs{\don,\smon,\blum}. The result
is an N=4 supersymmetric quantum mechanics based on the moduli space
${\cal M}^1$.

Without an explicit solution on $R^3 \times S^1$ with non-trivial
Wilson line it is difficult to calculate the metric on ${\cal M}^1$
directly. However it seems possible to determine the metric by
an indirect argument. First of all, note that the metric \mmet\ has
obvious Killing vectors which it inherits from translation symmetry in $R^3
\times S^1$
and from rotational symmetry in $R^3$.  Furthermore,  the three complex
structures
\comstr\ transform  as a triplet under the $SO(3)$ isometry.
The translation zero modes about the solution \two,\singlem\
are given by
\eqn\expzm{
\delta_\mu A_\nu = \p_\mu A^0_\nu-D^0_\nu A^0_\mu=F^0_{\mu \nu}
}
and obviously satisfy \bkgnd.
Substitution into \mmet\ gives the flat metric on
$R^3\times S^1 $.  As the remaining zero modes are independent of
the translation zero modes ${\cal M}^1$ must have the form
\eqn\mform{{\cal M}^1 = R^3 \times S^1 \times {\tilde{\cal M}}^1 }
with ${\tilde {\cal M}}^1$ a four-dimensional hyperK\"ahler manifold
with a $SO(3)$  isometry which rotates the hyperK\"ahler structure.
Following the arguments of \ah\ this requires that  ${\tilde {\cal M}}^1$ is
either the Atiyah-Hitchin metric or  the flat metric on
$R^4/Z_2$.  But we know that the metric \mmet\ which is inherited from
the gauge field kinetic term has a singularity as $\rho \rightarrow 0$
(ignoring possible string and gravitational corrections) so
${\tilde{\cal M}}^1$ must be $R^4/Z_2$ with the flat metric.

This argument can be checked by explicit calculation for $SU(2)$
instantons on $R^4$ \modrefs\ or on $R^3 \times S^1$ with vanishing
holonomy \unpub\
and one finds the expected result.

\subsec{Infrared Properties}

Having determined the moduli space we should analyze the spectrum of states
and compare with $S$-duality. Before doing this in this
subsection we will first make a few comments concerning the physical
interpretation
of the result for $ {\tilde{\cal M}}^1$. The most striking feature of
$ {\tilde{\cal M}}^1$ is that it is non-compact and hence the
quantum-mechanical
spectrum will be continuous. Of course the part of the moduli space
coming from translations, $R^3 \times S^1$,  is also non-compact,
but the resulting continuous spectrum  is just the usual continuum of momentum
states associated with translation invariance of the underlying theory.
The continuous spectrum on ${\tilde{\cal M}}^1$ means that there is
essentially an additional fake ``four-momentum'' which labels the
monopole configurations, at least in the moduli space approximation.
What is the physical origin of this effect? The non-compact direction
in $ {\tilde {\cal M}}^1$ is parametrized by the instanton scale size.
In conventional non-abelian Yang-Mills theory the effective coupling
will grow at large scale size, invalidating the semi-classical approximation.
In the $N=4$ theory considered here this is not the case since the
beta function vanishes identically \nfourref.

One may be tempted to think that the non-compactness arises
from very large configurations.
However this is not correct. It is known from studies of finite
temperature instantons \gpy\ and may be verified from the
solution \singlem\ that the effective size of an $H$ monopole is of
order $\rho/(1 + \rho^2/12 R^2)^{1/2}$ and is thus never much
larger than the radius $R$ of the $S^1$. Here the size is defined
by the falloff of the action or topological charge density of the
instanton. Thus
a ``plane wave'' configuration in ${\hat{\cal M}}^1$ is a configuration
in which the scale size $\rho$ varies from $0$ to $\infty$, but the
physical scale size is never large compared to $R$. It is also tempting
to think that the non-compactness is caused by coupling to massless
non-abelian gauge fields but this is also incorrect. The argument above
shows that ${\hat{\cal M}}^1$ is non-compact even with $SU(2)$ broken
to $U(1)$.
Thus the origin of the additional continuum in the $H$ monopole spectrum
must be due
to the coupling of the monopole
to the photon and massless scalars of the $N=4$ theory.
In principle it seems that this phenomenon could occur in any theory
with massless fields, but in most examples, including BPS monopoles
in $N=4$ Yang-Mills theory, it does not. This suggests that there
may be some problem in the correct identification of collective
coordinates in special theories with massless fields \foot{We thank N. Seiberg
for this suggestion and for discussions on this section.}.

One could try to address this issue by putting the system
in a box with twisted boundary conditions which remove the gauge
field and scalar zero modes following similar treatments in \thooft\
and \wittwist.  An immediate problem that arises
in trying to study $S$-duality in this context is that
neither electric charges nor magnetic charges exist on a compact
space. For electric charges this follow from Gauss's law
and for $H$ magnetic charges it follows from $dH = \alpha' {\rm Tr} F
\wedge F$. Integrating this equation over $T^4$ shows that the
instanton number, $p_1(V)$, must be trivial where $V$  is the $E_8\times E_8$
bundle.

Although we have little definite to say about $S$-duality
predictions on $T^4$, there are reasons for thinking this is a
direction worth pursuing.  These include connections with
the enhanced symmetries of string theory found in \moore,
the results on $N=4$ Yang-Mills theory on $T^4$ found
in \gir\ and results on the instanton moduli space on $T^4$.
For gauge group $SU(2)/Z_2$ it is known that the moduli space
of single instantons on $T^4$ with boundary conditions that remove the
gauge field zero mode is given by
\eqn\tfourmod{ {T^4 \times {\tilde M} \over Z_2 \times Z_2} }
where ${\tilde M}$ is an orbifold limit of a $K3$ manifold
\refs{\bmt, \kmir}.

We find the connection with $K3$ intriguing in that in the
infinite volume limit
the identification of ${\tilde {\cal M}}^1$ with $K3$ would naturally
explain the factor of $24$ degeneracy in the spectrum
of $H$ monopoles since $K3$ has precisely
$24$ harmonic forms.  On the other hand $K3$ is incompatible
with the requirement of $SO(3)$ holonomy and thus with the required
quantum numbers and we have argued that in fact
${\tilde {\cal M}}^1 = R^4/Z_2$. For a smooth, four-dimensional,
hyperk\"ahler manifold it is impossible to reconcile
the requirement of $SO(3)$ isometry with the requirement that
${\tilde {\cal M}}^1$ have $24$ harmonic forms.
In the following section we will argue that a reconciliation
may be possible for singular hyperk\"ahler manifolds
in string theory in that
 $R^4/Z_2$ clearly has $SO(3)$ isometry and also
has $24$ harmonic forms when the
forms are counted using ideas from orbifolds.

\subsec{Quantization on ${\hat{\cal M}}^1$}

In the hope that the infrared problems can be resolved let us now
discuss
a direct treatment of the quantization in infinite volume.
The first problem we encounter in thinking about quantization on
${\cal M}^1$ is the breakdown of the low-energy field theory approximation.
Although we can work in a region of the moduli space of Narain
compactifications where $H$ monopoles are light, as
$\rho \rightarrow 0$ the field strength becomes more and more concentrated
at the center of the instanton and eventually is large compared to
the string scale or Planck scale. It thus seems likely that gravitational
and string corrections to the moduli space will become important for
scale sizes small compared to $\sqrt{\alpha'}$. On the other hand,
the moduli space is tightly constrained by the demands of $N=4$ supersymmetry
and rotational invariance. We can imagine four general possibilities.

First, we can imagine that there are no $\alpha'$ corrections to the
moduli space ${\hat{\cal M}}^1$ and that the spectrum is given
by the spectrum of $N=4$ supersymmetric quantum mechanics on
${\cal M}^1$. In this case we find  $16 \times 8$ states with the
factor of $16$ arising as usual from harmonic forms on
$R^3 \times S^1$ and the factor of $8$ arising from the $8$ $Z_2$
invariant forms on $R^4/Z^2$.
This spectrum is not
in agreement with that predicted by $S$-duality.

Second, it is possible that gravitational and other $\alpha'$ corrections
change the moduli space at short distances. One natural guess is
that the orbifold singularity at the origin is resolved as in the
construction of $K3$ from the $T^4/Z_2$ orbifold, that is by replacing
$R^4/Z_2$ by the Eguchi-Hanson metric. However on closer inspection
this seems unlikely. We argued above that the moduli space should
be hyperK\"ahler with an $SO(3)$ isometry rotating the three complex
structures.
The Eguchi-Hanson metric is hyperK\"ahler and has $SO(3)$ isometry, but
the three complex structures are left invariant by the $SO(3)$ action \foot{
$R^4$  has two sets of commuting hyperK\"ahler structures given
in terms of the 't Hooft symbols by $J^{+a}_{\mu \nu} = - \eta^a_{\mu \nu}$,
$J^{-a}_{\mu \nu} = - {
\bar \eta} ^a_{\mu \nu} $. Under the $SO(4)=SU(2)_+ \times SU(2)_-$ isometry
these
transform as $(3,1)$ +$(1,3)$. The $SO(3)$ of rotations we are interested
in is the diagonal of $SU(2)_+$ and $SU(2)_-$ and both sets of
complex structures transform
as  triplets.  The Eguchi-Hanson metric has a self-dual $SU(2)$ isometry and
an anti-self-dual hyperK\"ahler structure which is thus invariant under
the isometry. See \gibb\ for more details.}

A third possibility is that the correct low-energy monopole dynamics is not
that of
standard $N=4$ supersymmetric quantum mechanics but rather involves
correction terms coming from the supergravity fields \foot{This suggestion
is due to E. Witten.}. In this case the
connection between states and harmonic forms may also be modified.
An example of such a modification is discussed in \rohmw.  This
possibility deserves further thought, but it seems unlikely to lead to
the required number of states.

A fourth possibility is that
the moduli space is in fact
$R^3 \times S^1 \times R^4 / Z_2$, but that the correct treatment
of collective coordinates in string theory differs from that
in low-energy field theory due to the orbifold singularity in
the moduli space. Since the treatment of collective coordinates
in string theory is for the most part unknown, we can only speculate as
to how string theory and particle theory might differ.
However with that warning there
is at least one plausible guess which turns out to be in agreement
with the predictions of $S$-duality. The guess is that the treatment of
collective coordinates in string theory involves promoting the
collective coordinates not just to quantum mechanical variables,
$A_\mu(x,Z^i) \rightarrow A_\mu(x,Z^i(t))$, but to two-dimensional
string variables, $Z^i \rightarrow Z^i(\tau, \sigma)$ in order to
be compatible with modular invariance. It seems unlikely that one
really wants the full spectrum of a new string coordinate, so there
must presumably be a truncation to just the low-lying spectrum, perhaps
as in topological field theory.
Thus our guess is that the low-lying spectrum of monopoles
with a moduli
space possessing orbifold singularities is given by
the low-lying spectrum of superstring theory on the moduli space. Note
that this does not affect the treatment of the BPS monopole moduli
space which is known in general to be non-singular.

Following this line of speculation we wish to calculate the
low-lying spectrum
of superstring theory on $R^3 \times S^1 \times R^4/Z_2$.
Since the
$R^3 \times S^1$ part will be the same in string theory and
particle theory, we can concentrate on the $R^4/Z_2$ factor.
The low-lying superstring spectrum on $R^4/Z_2$ can be computed
using orbifold techniques. The partition function is
\eqn\zorb{Z = \half (Z_{1,1} + Z_{1, \theta} + Z_{\theta,1} +
                Z_{\theta,\theta} )}
where $\theta$ is the $Z_2$ transformation and as usual $Z_{g,h}$
denotes the partition function for strings with boundary conditions
twisted by $g$ in the $\sigma$direction and by $h$ in the $\tau$ direction.
The first two terms in \zorb\ give the point particle
spectrum projected onto invariant states. The low-lying states from
this sector are thus $8$-fold degenerate. The last two terms in
\zorb\ comes from twisted strings which close only up to a $Z_2$
transformation. They can be determined by modular invariance from
$Z_{1,\theta}$. However since $Z_{1,\theta}$ is independent of momenta,
it is the same on $R^4/Z_2$ as on $T^4/Z_2$. Thus the spectrum from
the twisted sector is the same as on $T^4/Z_2$, that is $16$ fold
degenerate. Thus the total degeneracy is $24$, in agreement with
$S$-duality. The basic point is that the degeneracy of $16$ in the
twisted sector is required by modular invariance whether we are
on $T^4/Z_2$ or on $R^4/Z_2$. On $T^4/Z_2$, which is an orbifold
limit of $K_3$, the factor of $16$ can be viewed as arising from
the $16$ fixed points of the $Z_2$ transformation. We might try to
define string propagation on $R^4/Z_2$ by taking the infinite radius
limit of $T^4/Z_2$, but this is problematic since if we take one of
the fixed points at the origin, the $15$ states at the other fixed points
move off to infinity in this limit. On the other hand, we don't see
any obstacle to directly defining string propagation on $R^4/Z_2$
with $16$ states localized at the one fixed point of the $Z_2$
transformation on $R^4$.

Finally we can ask if this assumption is also compatible with the
quantum numbers required by $S$-duality. Recall that this
requires that the $24$ states arising from ${\tilde {\cal M}}^1$ transform
as $16 (1,1) + (5,1)+(1,3)$ under the $SO(5) \times SO(3)$ unbroken
symmetry.  For a charge one instanton it is known that all fermion
zero modes can be obtained by supersymmetry transformations.
Thus, before modding out $R^4$ by $Z_2$ the fermion zero modes
transform as $(4,2)$ as do the supersymmetry transformations.
Quantization of the zero modes gives
the representations $ (5,1)+(1,3)+(4,2)$ with the states $(5,1) +(1,3)$
created from the vacuum by the
action of an even number of fermion creation operators (even cohomology)
and the states $(4,2)$ created by the action of an odd number of fermion
creation operators (odd cohomology). The $Z_2$ acts
as $-1$ on the fermion zero modes, so projecting onto
$Z_2$ invariant states we get $(5,1)+(1,3)$. In addition there
are $16$ massless states arising in the ``twisted sector''. The massless states
arise in a sector with anti-periodic fermions and anti-periodic
bosons and are thus singlets of the fermion current algebra.  We
thus  presume that these states
are singlets under $SO(5) \times SO(3)$. The tensor product with
the $16$ states coming from $R^3 \times S^1$  gives the spectrum
required by $S$-duality, including massive spin two magnetic
monopoles.

\newsec{Conclusions}

It is clear that we have not presented unambiguous new evidence
for $S$-duality in string theory. However,
thinking about the structures that would be required
by $S$-duality has led to some specific suggestions about the
structure of string theory. In particular, the question of the
treatment of orbifold singularities in the moduli space of solutions
to string theory may give an example where the treatment of
collective coordinates in string theory differs in a qualitative
way from the treatment in low-energy field theory.  It also
seems  that the moduli spaces encountered in treating various
kinds of magnetic monopoles are quite rigid. It may well be
that there is a non-renormalization theorem for  monopole
moduli spaces which states that they receive no higher order
corrections in $\alpha'$. This would be consistent with our conjecture
regarding $R^4/Z_2$.

If our guess as to the treatment of orbifold singularities
is correct then we  would have new evidence for the existence
of  $S$-duality in heterotic string theory.  Furthermore, our results would
provide evidence for this duality not just in $N=4$ Yang-Mills theory
but also in low-energy string theory on $T^6$.
The $H$-monopoles we have described
depend in general on the compactified dimensions and cannot be
viewed purely as four-dimensional solutions. We should also
reiterate that we have probed only the first of many predictions
that follow from $S$-duality in string theory. In light of the conjecture
made in \senb\ it will be particularly interesting to investigate whether
similar features appear in the multi-$H$-monopole moduli space and
to extend these ideas to the infinite tower of monopoles required
by $S$-duality. These excitations include states
of arbitrarily high angular momentum and they are intrinsically stringy,
that is they do not have a Kaluza-Klein origin.

In $N=4$ Yang-Mills theory
the monopole multiplet contains massive spin one monopoles. As the only
consistent formulation of fundamental spin one objects that we know of
is spontaneously broken gauge theory, it seems likely that there
is a dual formulation in which the monopoles become gauge bosons.
Similarly, the existence of massive spin two monopoles in toroidal
compactifications of low-energy string theory suggest the existence of a
dual formulation involving extended objects such as strings. Note
that the appearance of  massive spin two $H$-monopoles does not
depend on our assumptions concerning collective coordinates
in string theory. Standard $N=4$ supersymmetric quantum mechanics
on $R^4/Z_2$ gives $8$ of the $24$ states required by $S$-duality
and these states when tensored with the states arising from $R^3 \times S^1$
include spin two monopoles.

Although we lack any hard evidence as yet, it would be extremely
interesting if there exists an extension
of $S$-duality beyond toroidal compactifications. Given the
suggestion of \sena\ that it should be thought of as a discrete
gauge symmetry makes this seem quite likely, although
the action is likely to be much less trivial in other
backgrounds.
Given that the monopoles
we have discussed are really extended fivebranes wrapped around
some internal dimensions, it is natural
to speculate that the dual theory should be a quantum theory of heterotic
fivebranes. Fivebranes occur as the natural objects which couple to
the dual formulation of $D=1$ supergravity \duff,  and explicit fivebranes
solutions of low-energy string theory  exist and have properties suggesting
such a dual formulation
\hetsol. In addition, the string states which are dual to the $H$-monopoles
discussed above have certain characteristics of solitons \dab\ that suggest
that they might arise as soliton solutions in a dual theory of fivebranes
\dufflutwo.

However there are severe difficulties in constructing a quantum theory
of fivebranes and it is widely believed that such a theory does
not exist, at least not with a spectrum which resembles that of string theory.
The fivebrane solutions of \hetsol\ are based on Yang-Mills
instantons. The results here suggest to us  that $S$-duality in string theory
may be more subtle than a duality involving fivebranes as extended
objects. It seems instead that the moduli space of   instantons
encodes certain spacetime structures in somewhat the same way that
Riemann surfaces and various structures on them encode spacetime
properties in string theory. In studying the single $H$ monopole
moduli space we found a moduli space which was locally
$M_{ss} \times M_{bs}$
with $M_{ss}$ the underlying four-manifold $R^3 \times S^1$
and $M_{bs}= R^4/Z_2$. Modulo
our assumption about the counting of harmonic forms in string theory,
$M_{ss}$ reflects the structure of the right-moving superstring ground
state in its cohomology in that  the even cohomology and odd
cohomology are both $8$ dimensional while $M_{bs}$ reflects the structure
of the left-moving string first excited state in having no odd
cohomology and $24$ dimensional even cohomology.
Perhaps we should be looking for a reformulation
of string theory in which the primary objects are  moduli spaces
of self-dual connections on four manifolds rather than moduli
spaces of Riemann surfaces.

\bigskip
\centerline{\bf Acknowledgements}\nobreak

We thank T. Banks, C. Callan, G. Gibbons, N. Hitchin,
E. Martinec, G. Moore, D. Morrison,  N. Seiberg,
A. Sen, S. Shenker, A. Strominger, C. Vafa and E. Witten for
many helpful discussions.
JH would like to acknowledge the
hospitality and support of the Rutgers Theory Group while this
work was in progress. JG and JH also thank the Newton Institute for
Mathematical Sciences for hospitality while this work was being completed.
This work was supported in part
by NSF
Grant No.~PHY90-00386.  JPG\ is supported by a grant from the
Mathematical Discipline Center of the Department of Mathematics,
University of Chicago. JH\ also acknowledges the support of NSF PYI
Grant No.~PHY-9196117.

\listrefs
\end